\documentclass[aps,prl,superscriptaddress,11pt]{revtex4-1}

\usepackage{graphicx}
\usepackage[T1]{fontenc} 
\usepackage{graphicx,xcolor}
\usepackage{amssymb,amsmath,amsthm}
\usepackage{mathrsfs}
\usepackage{amsfonts}
\usepackage{booktabs}
\usepackage{siunitx}
\usepackage{booktabs}
\usepackage{natbib}

\begin{document}


\title{Coherent control of the orbital occupation driving the insulator-to-metal Mott transition in V$_2$O$_3$}

\author{Paolo Franceschini}
\email[]{paolo.franceschini@unicatt.it}
\affiliation{Department of Mathematics and Physics, Università Cattolica del Sacro Cuore, IT-25133 Brescia, Italy}
\affiliation{ILAMP (Interdisciplinary Laboratories for Advanced
Materials Physics), Università Cattolica del Sacro Cuore, IT-25133 Brescia, Italy}
\affiliation{Department of Physics and Astronomy, KU Leuven, B-3001 Leuven, Belgium}
\affiliation{Present address: CNR-INO (National Institute of Optics), via Branze 45, IT-25123 Brescia, Italy}

\author{Veronica R. Policht}
\affiliation{Department of Physics, Politecnico di Milano, IT-20133 Milano, Italy}

\author{Alessandra Milloch}
\affiliation{Department of Mathematics and Physics, Università Cattolica del Sacro Cuore, IT-25133 Brescia, Italy}
\affiliation{ILAMP (Interdisciplinary Laboratories for Advanced
Materials Physics), Università Cattolica del Sacro Cuore, IT-25133 Brescia, Italy}
\affiliation{Department of Physics and Astronomy, KU Leuven, B-3001 Leuven, Belgium}

\author{Andrea Ronchi}
\affiliation{Department of Mathematics and Physics, Università Cattolica del Sacro Cuore, IT-25133 Brescia, Italy}
\affiliation{ILAMP (Interdisciplinary Laboratories for Advanced
Materials Physics), Università Cattolica del Sacro Cuore, IT-25133 Brescia, Italy}
\affiliation{Department of Physics and Astronomy, KU Leuven, B-3001 Leuven, Belgium}
\affiliation{Present address: Pirelli Tyre S.p.A, viale Piero e Alberto Pirelli 25, IT-20126 MIlano, Italy}

\author{Selene Mor}
\affiliation{Department of Mathematics and Physics, Università Cattolica del Sacro Cuore, IT-25133 Brescia, Italy}
\affiliation{ILAMP (Interdisciplinary Laboratories for Advanced
Materials Physics), Università Cattolica del Sacro Cuore, IT-25133 Brescia, Italy}

\author{Simon Mellaerts}
\affiliation{Department of Physics and Astronomy, KU Leuven, B-3001 Leuven, Belgium}

\author{Wei-Fan Hsu}
\affiliation{Department of Physics and Astronomy, KU Leuven, B-3001 Leuven, Belgium}

\author{Stefania Pagliara}
\affiliation{Department of Mathematics and Physics, Università Cattolica del Sacro Cuore, IT-25133  Brescia, Italy}
\affiliation{ILAMP (Interdisciplinary Laboratories for Advanced
Materials Physics), Università Cattolica del Sacro Cuore, IT-25133 Brescia, Italy}

\author{Gabriele Ferrini}
\affiliation{Department of Mathematics and Physics, Università Cattolica del Sacro Cuore, IT-25133  Brescia, Italy}
\affiliation{ILAMP (Interdisciplinary Laboratories for Advanced
Materials Physics), Università Cattolica del Sacro Cuore, IT-25133 Brescia, Italy}

\author{Francesco Banfi}
\affiliation{FemtoNanoOptics group, Université de Lyon, CNRS, Université Claude Bernard Lyon 1, Institut Lumière Matière, F-69622 Villeurbanne, France}

\author{Michele Fabrizio}
\affiliation{Scuola Internazionale Superiore di Studi Avanzati (SISSA), IT-34136 Trieste, Italy}

\author{Mariela Menghini}
\affiliation{IMDEA-Nanociencia, E-28049 Madrid, Spain}

\author{Jean-Pierre Locquet}
\affiliation{Department of Physics and Astronomy, KU Leuven, B-3001 Leuven, Belgium}

\author{Stefano Dal Conte}
\affiliation{Department of Physics, Politecnico di Milano, IT-20133 Milano, Italy}

\author{Giulio Cerullo}
\affiliation{Department of Physics, Politecnico di Milano, IT-20133 Milano, Italy}

\author{Claudio Giannetti}
\email[]{claudio.giannetti@unicatt.it}
\affiliation{Department of Mathematics and Physics, Università Cattolica del Sacro Cuore, IT-25133  Brescia, Italy}
\affiliation{ILAMP (Interdisciplinary Laboratories for Advanced
Materials Physics), Università Cattolica del Sacro Cuore, IT-25133  Brescia, Italy}


\date{\today}

\begin{abstract}
Managing light-matter interactions on timescales faster than the loss of electronic coherence is key for achieving full quantum control of the final products in solid-solid transformations. In this work, we demonstrate coherent electronic control of the photoinduced insulator-to-metal transition in the prototypical Mott insulator V$_2$O$_3$. Selective excitation of a specific interband transition with two phase-locked light pulses manipulates the orbital occupation of the correlated bands in a way that depends on the coherent evolution of the photoinduced superposition of states. A comparison between experimental results and numerical solutions of the optical Bloch equations provides an electronic coherence time on the order of 5 fs. Temperature-dependent experiments suggest that the electronic coherence time is enhanced in the vicinity of the insulator-to-metal transition critical temperature, thus highlighting the role of fluctuations in determining the electronic coherence. These results open new routes to selectively switch the functionalities of quantum materials and coherently control solid-solid electronic transformations.   

\end{abstract}


\maketitle
\clearpage

\section{Introduction}

The ability to control matter transformations along quantum coherent pathways is key for opening new frontiers in condensed matter physics, with a broader impact on the development of novel quantum technologies \cite{Basov2017,Koshihara2022}. In contrast with conventional state transitions, in which electrons can be considered as incoherent degrees of freedom instantaneously coupled to external reservoirs (phonons, vibrations, and charge or spin excitations), coherent control protocols are based on the creation of a quantum coherent superposition of states that freely evolve and determine the output of the transformation before any decoherence process takes place \cite{SHAPIRO2000287,Zewail2000}. 

Early efforts to achieve optical coherent control exploited the long coherence times ($T_2 \gtrsim$1 ps) of atomic and molecular systems to control the output of specific chemical reactions \cite{Zewail1988,Zewail1989,Potter1992,Zewail2000,Assion1998,Herek2002}. The solid state counterpart, i.e., the control of the output of a thermodynamic phase transition, is far more challenging given decoherence timescales on the order of a few femtoseconds due to the extremely efficient coupling of charge excitations to the environment. Recent attempts at coherent control in solid state systems have demonstrated optical control of the insulator-to-metal transition in organic correlated crystals \cite{Matsubara2014} and indium wires \cite{Horstmann2020} by exploiting the relatively long-lived vibrational coherence. In these studies, a combination of two short phase-coherent light pulses was shown to enable the switching of the structural phase of the system in a way that depends on the instantaneous nuclear position during the oscillation of the structural amplitude modes connecting the two phases \cite{Horstmann2020}. Coherent schemes have additionally been implemented to control the dynamics of photoemitted electrons in metals and superconductors \cite{PetekPSS1997,OgawaPRL78_1997,PetekPRL79_1997,NesslerPRL81_1998} and the instantaneous electronic photo-currents in semiconductors \cite{AtanasovPRL1996,HachePRL1997}, without driving any phase transformation in the material. To date, managing the coherent dynamics of electronic states that control the output of a solid-solid phase transition is still an unexplored field.  

\begin{figure*}[t]
\includegraphics[keepaspectratio,clip,width=\textwidth]{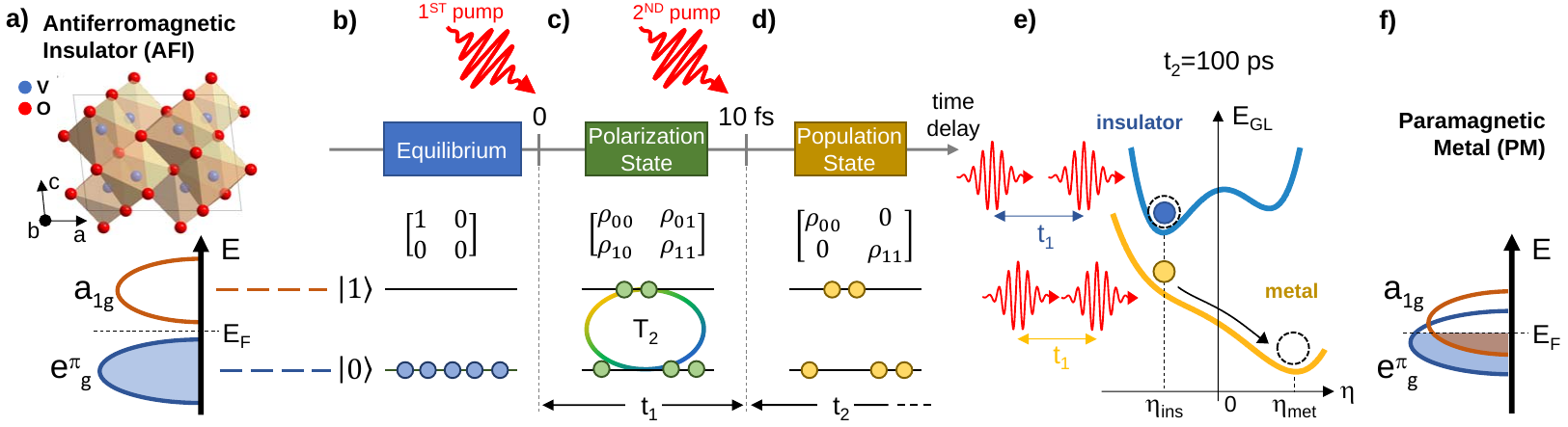}
\caption{\label{fig:cartoon}{\bfseries Cartoon of the Experiment.} a) Crystal structure and sketch of the electronic band structure near the Fermi level ($E_F$) for the antiferromagnetic insulating (AFI) phase of V$_2$O$_3$. b-d) Two-level system representing the orbital population excited by two coherent pump pulses and e) free-energy diagram {($E_{GL}$, with $\eta$ being the order parameter)} showing the phase of the system {(insulating or metallic at $\eta_{\scriptsize{\mbox{ins}}}$ or $\eta_{\scriptsize{\mbox{met}}}$, respectively)}. The ground and excited many-body states, characterized by different occupations of the the $e^{\pi}_g$ and $a_{1g}$ electronic levels, are identified as $\left|0 \rangle \right.$ and $\left|1 \rangle \right.$. b) When the first pump pulse arrives, the system is at equilibrium (insulating phase), i.e., in the ground state $\left|0 \rangle \right.$. c) After the first excitation, the system is left in a \emph{polarization state},  described by non-zero off-diagonal terms ($\rho_{01}$ and $\rho_{10}$) of the density matrix. The second phase-coherent pulse manipulates the coherent superposition of levels $\left|0 \rangle \right.$ and $\left|1 \rangle \right.$ within the coherence time $T_2$. After the coherent photo-excitation, the system is left in a population state d), with zero off-diagonal terms, which evolves towards a thermal state. {e)} The $t_1$-dependent population determines the energy potential profile and, consequently, the final state of the system at large delay time, $t_2=\SI{100}{ps}$. f) Electronic band structure near the Fermi level ($E_F$) for the paramagnetic metallic (PM) phase of V$_2$O$_3$. Crystal structure in a) adapted from Ref. \cite{Rozier2002}, and electronic band structure in a) and f) adapted from Ref. \cite{Qazilbash2008}.}
\end{figure*}

Here, we report evidence of the electronic coherence of the interband transitions relevant for the insulator-to-metal transition (IMT) of V$_2$O$_3$ by means of optical manipulation with two extremely short phase-locked light pulses. V$_2$O$_3$ undergoes a phase transition at $T_{IMT} \sim \SI{180}{K}$ from a low-temperature antiferromagnetic monoclinic insulator (AFI) to a high-temperature paramagnetic corundum metal (PM). The combination of strong on-site Coulomb repulsion, crystal field splitting, and trigonal distortion of the V-O octahedra gives rise to a manifold of electronic states within the vanadium 3$d$ levels. In particular, the conductive state of V$_2$O$_3$ is mainly determined by the occupation of the lowest levels, i.e. the $e^{\pi}_g$ doublet, mainly oriented in the $a$-$b$ plane, and the upper $a_{1g}$ singlet, mainly oriented along the $c$ axis (see Fig. \ref{fig:cartoon}a). At $T_{IMT}$, the shortening of the V-V dimers along the hexagonal $c$ axis favors an increase of the $a_{1g}$ occupation ($n_{a_{1g}}$) at the expense of the $e^{\pi}_g$ one ($n_{e^{\pi}_g}$) and drives the transition from a Mott insulator to a metal \cite{Poteryaev2007,Park2000,Ronchi2021}. The $a_{1g}$ orbital occupation, though being finite in both phases \cite{Goodenough1971ARMS,Goodenough1971PSSC,Park2000}, jumps upwards at the transition and thus can be considered as its control parameter. 

Optical transitions in the visible region can be used to non-thermally manipulate $n_{a_{1g}}$. Visible light pulses have been demonstrated to trigger the transformation from an insulator to a metastable metal by the suitable excitation of interband transitions  \cite{LiuPRL2011,SandriPRB2015,LantzNC2017,Ronchi2019,Ronchi2021}. The underlying concept of the coherent control protocol is based on the excitation of the $e^{\pi}_g\rightarrow a_{1g}$ optical transition by two phase-coherent pulses (Fig. \ref{fig:cartoon}b-1e). For simplicity, we will refer to the insulating many-body ground state as $\left| 0\rangle \right.$ and to the excited state as $\left| 1\rangle \right.$.  The first pump pulse excites the insulating ground state, characterized by the equilibrium occupation $n_{a_{1g}}$, and creates a quantum superposition of states $\left| \Psi\rangle \right.$=$p_0$ $\left| 0\rangle \right.$+$p_1$ $\left| 1\rangle \right.$, where $|p_{0(1)}|^2$ is the time-dependent probability of finding the system in the state $\left| 0(1)\rangle \right.$ (Fig. \ref{fig:cartoon}c). If we consider the density matrix $\widehat{\rho}$ $=$ $\left| \Psi\rangle \right. \left. \langle \Psi \right|$, the quantum polarization state generated by the first pump pulse is described by non-zero off-diagonal terms, $\rho_{nm}$, which are eventually destroyed by the decoherence brought by the environment, here composed by phonons, and spin and charge excitations. The following interaction with the second phase-locked pump pulse leaves the system in a population state characterized by increased $n_{a_{1g}}$, encoded in the diagonal population term ($\rho_{11}$) of the density matrix (Fig. \ref{fig:cartoon}d). The orbital population variation induced by the excitation protocol thus depends on the instantaneous polarization state at the time of the interaction with the second pump pulse. 
{As demonstrated in Ref. \citenum{Ronchi2021}, the IMT can be described via a Landau-Ginzburg energy functional, $E_{GL}$ (Fig. \ref{fig:cartoon}e), which takes the form:
\begin{equation}
E_{GL}(n_{a_{1g}})\propto{\left[\eta-\eta_{\scriptsize{\mbox{met}}}\right]^2\left[\eta-\eta_{\scriptsize{\mbox{ins}}}\right]^2-g(n_{a_{1g}} ,T)\eta}
\label{eq_freeenergy}
\end{equation}
where $\eta$ is the order parameter assuming the values $\eta_{\scriptsize{\mbox{met}}}>0$ and  $\eta_{\scriptsize{\mbox{ins}}}<0$ in the metallic and insulating phases, respectively. The effect of the interaction with the light pulses is accounted for by the coupling term $g(n_{a_{1g}},T)$, which is negative at equilibrium for $T<T_{IMT}$ and becomes positive, thus stabilizing the metallic phase, when either $n_{a_{1g}}$ is increased by the pump pulses or $T>T_{IMT}$.
As a consequence, by preparing the $\left| \Psi\rangle \right.$ many-body state we can coherently control the final population difference $n_{a_{1g}}-n_{e^{\pi}_g}$=$\rho_{11}-\rho_{00}$ and, in turn, the free energy of the system. The combined action of the two phase-coherent pump pulses thus leaves the system in a nonequilibrium configuration, which evolves towards a metastable metallic state, characterized by $\eta_{\scriptsize{\mbox{met}}}$, within the timescale ($\sim$50 ps) necessary to complete the electronic and structural transformation \cite{Ronchi2019,Ronchi2021}.}

In conventional incoherent excitation schemes, the amount of material undergoing the phase transition is strictly proportional to the intensity of the excitation light. In contrast, the coherent dynamics addressed by the two-pump coherent experiment proposed here results in multiple effects. First, if the two-level system is excited by two phase-coherent pulses with photon energy $\hbar \omega_p$=$\hbar \omega_{01}$+$\delta \omega$, where $\hbar \omega_{01}$ is the energy difference between the two levels and $\delta \omega$ the energy detuning, the oscillation of $\rho_{11}-\rho_{00}$ as a function of the delay between the two coherent pumps is pinned to $\omega_{01}$, in the limit of large $T_2$. This leads to a detectable frequency difference between the time-domain linear interferogram of the two pump pulses and the signal related to the photoinduced IMT. Second, if the dephasing time of the electronic coherence state generated by the first pulse is comparable to or longer than the pulse-width, coherence effects can be observed beyond the strict temporal overlap of the two pump pulses. In particular, oscillations of the final population can survive longer than the time-domain linear interferogram of the two pump pulses, thus leading to a variation $\delta \gamma=\gamma_s-\gamma_p$ of the spectral width of the signal ($\gamma_s$) relative to the spectral width of the pump pulse ($\gamma_p$), which is related to the IMT.

In order to resolve possible signatures of the coherent dynamics described above, it is crucial to maintain a high degree of phase coherence between the two pump pulses and to accurately tune the experimental parameters so as to maximize the effects. This experiment requires temporally short and spectrally broadband light pulses whose spectral widths cover the spectral region of the expected frequency shift, while providing enough excess energy to overcome the insulator-to-metal transformation barrier (see Sec. S2 in the Supplementary Information, SI). The choice of the experimental parameters was informed by the numerical solution to the Optical Bloch Equations (OBE) (see Sec. S3 in the SI), which allows us to simulate the dynamics of the $\rho_{11}$ population in the presence of an effective decoherence driven by the coupling of the electronic wave functions with the environment. Considering realistic pulse durations ($\sim \SI{30}{fs}$) and reasonable tunability ($\sim \SI{50}{meV}$, see Fig. S3 in the SI) around $\omega_{01}$, signatures of coherent effects are observable for $T_2$ as small as few femtoseconds, provided a phase stability of the order of 1/1000 of the optical cycle is achieved.

In the present work, the IMT is optically triggered by means of two phase-coherent pump pulses (with duration $<\SI{28}{fs}$ as measured by polarization-gated frequency-resolved optical gating, PG-FROG) generated by translating-wedge-based identical-pulses-encoding system (TWINS) technology, a collinear interferometer based on birefringent wedges capable of tuning the relative delay ($t_1$) between the two pulse replicas with attosecond precision and excellent phase stability ($\sim\SI{2}{as}$, see Sec. S4 in the SI for more details) \cite{Brida2012,RevSciInstr2014,Oriana2016,Preda2017}. A third optical pulse arriving at fixed delay $t_2$ after the excitation pulses probes the final state of the system, thus providing a $t_1$-dependent pump-induced relative reflectivity variation, $\delta R/R$ ($t_1, t_2$) (see Sec. S5 in the SI for a detailed description of the experimental setup). In this work, we focus on the interband electronic transition at $\sim \SI{2.4}{eV}$ (Fig. S1 in the SI), which is a transition between the $e^{\pi}_g$ and $a_{1g}$ bands and is particularly sensitive to the insulator-to-metal transformation \cite{Ronchi2019}. The sample under study consists of a 50-nm-thick V$_2$O$_3$ thin film deposited by oxygen assisted molecular beam epitaxy (MBE) on a (0001)-oriented sapphire (Al$_2$O$_3$) substrate, with the $c$ axis perpendicular to the surface \cite{Dillemans2014}. 

\begin{figure}[t]
\includegraphics[keepaspectratio,clip,width=0.45\textwidth]{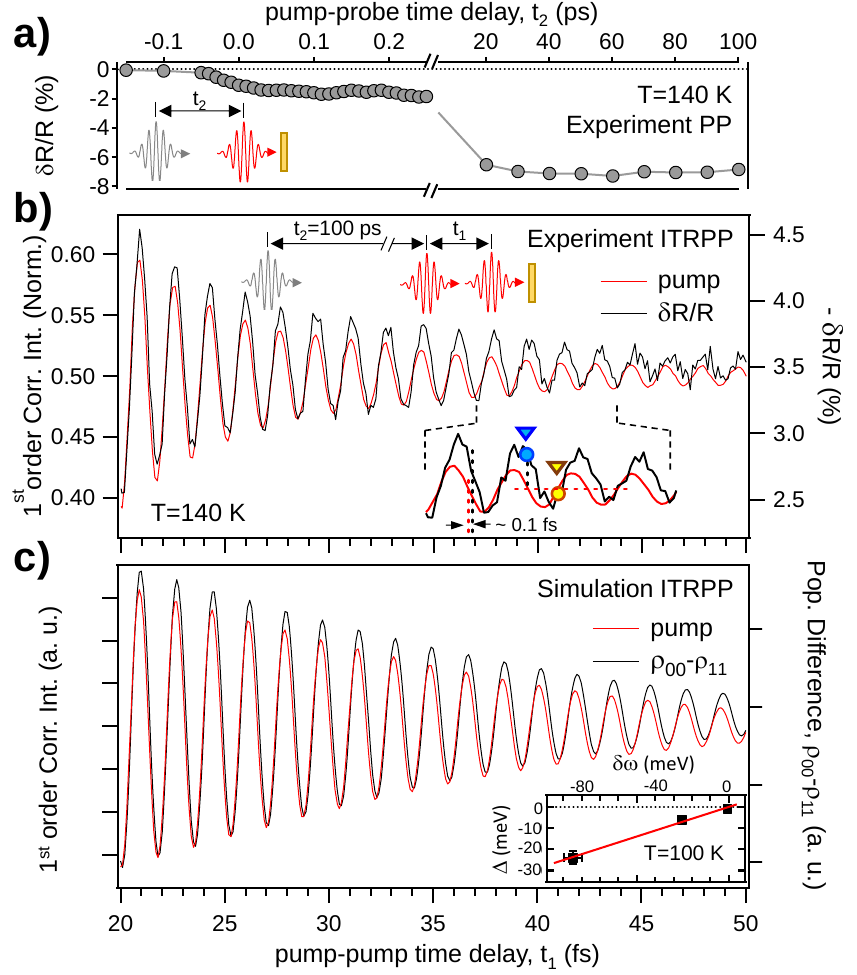}%
\caption{\label{fig:interferograms} {\bfseries Interferometric time-resolved pump-probe (ITRPP).} a) Relative reflectivity variation $\delta R/R \left(t_2 \right)$ as a function of $t_2$ detected at $\SI{2.42}{eV}$ photon energy (standard pump-probe, PP, experiment as sketched in the inset). b) Relative reflectivity variation $\delta R/R \! \left(t_1 \right)$ detected at $\SI{2.42}{eV}$ photon energy as a function of the pump-pump delay $t_1$ (black line), at a fixed probe delay $t_2=\SI{100}{ps}$, compared with the first-order correlation function of the pump pulse (red line). The experimental data in panels a) and b) have been obtained at $T=\SI{140}{}$ K with an incident pump fluence of $0.4$ mJ/cm$^2$. {Inset: Expanded view highlighting the different $\delta R/R$ value (markers) at $t_1$ delay times corresponding to the same pump intensity: $t_1$=39.8 fs (blue marker) and $t_1$=40.7 fs (yellow
marker).} c) Calculated two-level population difference as a function of the pump-pump delay $t_1$ (black line) compared with the first-order correlation function of the pump pulse (red line). Inset (panel {c}): Excitation-energy-dependent shift at fixed temperature $T=100$ K. A linear model (red line) has been fitted to the experimental data (black markers). The simulation has been performed by solving OBE with $T_2=\SI{5}{fs}$ and $\delta \omega=\SI{0.06}{eV}$. Signals in b) and c) are shown for $t_1\geq \SI{20}{fs}$ to highlight the frequency shift.}
\end{figure}

We first observe the dynamics of the incoherent IMT by performing a conventional pump-probe experiment, corresponding to the case of $t_1=0$, at $T= \SI{140}{K}$. For fluences higher than $\sim$4 mJ/cm$^2$, the $\delta R/R (t_2)$ signal at long delays ($t_2>50$ ps) shows the same reflectivity decrease observed during the thermally driven IMT \cite{Ronchi2021}. For intermediate fluences (<0.5 mJ/cm$^2$), as those used for the trace reported in Fig. \ref{fig:interferograms}a and in the following experiments, the signal is linearly proportional to the pump excitation and to the metallic filling fraction variation \cite{Ronchi2019}.
The relative reflectivity variation dynamic is characterized by a build-up of the order of few tens of ps corresponding to the photoinduced nucleation and growth of metallic domains \cite{Ronchi2019,Abreu2015}. At $t_2=\SI{100}{ps}$, the phase transformation is complete as indicated by the $t_2$-dependent signal (Fig. \ref{fig:interferograms}a). Therefore this $t_2$ value is chosen as the detection time for the interferometric experiment, in which the transient $\delta R/R (t_1)$ signal is recorded as a function of $t_1$. The $t_1$-dependent relative reflectivity variation signal is reported in Fig. \ref{fig:interferograms}b. The $\delta R/R (t_1)$ signal (black solid line) is shown together with the linear interferogram (first-order correlation function) of the pump pulse pair (red solid line). The oscillatory pattern of $\delta R/R$ shows a progressive dephasing with respect to the pump interferogram thus suggesting a frequency shift between the two signals. The phase difference accumulated after 25 oscillations corresponds to a delay of $\sim\SI{100}{as}$, which is well above the phase stability of our setup. {The observed dephasing indicates that the correlated $\delta R/R (t_1)$ signal, proportional to the amount of material driven into the metallic phase, is no longer proportional to the pump intensity. The inset of Fig. \ref{fig:interferograms}b shows that, given the same pump intensity for example at $t_1=\SI{39.8}{fs}$ and $t_1=\SI{40.7}{fs}$, $\delta R/R (t_1)$ after 100 ps assumes two values differing by about 7 \%.}
The frequency shift between the two interferograms in Fig. \ref{fig:interferograms}b is evaluated by comparing the central frequencies of the spectra obtained by Fourier transforming the time-domain data with respect to $t_1$. Details of the analysis process can be found in Sec. S5 of the SI. We find that the frequency of the $\delta R/R (t_1)$ signal is red-shifted by an amount corresponding to $\Delta=\left( \SI{4}{} \pm \SI{1}{} \right)$ meV with respect to the pump linear interferogram. At the same time, oscillations of the $\delta R/R (t_1)$ signal persists for a longer time interval than the temporal width of the pump interferogram, as indicated by the spectral width variation $\delta \gamma/\gamma_p$=$-\left( \SI{2.6}{} \pm \SI{0.3}{} \right) \, \%$. From the numerical solution of OBE, we pinpoint that the energy shift $\Delta$ depends on the coherence time and that it is smaller than the detuning, $\left| \Delta \right| < \left| \delta \omega \right|$, for finite $T_2$. In the limit $T_2\rightarrow \infty$, $\Delta$ coincides with $\delta \omega$ (see Sec. S2 in the SI for more details).

The experimental data shown in Fig. \ref{fig:interferograms} are compared to the results of the OBE simulations, which are used to compute the time-dependent population difference, $\rho_{00}-\rho_{11}  \left( t_1 \right)$, following the interaction with the two phase-coherent pump pulses (black solid line in Fig. \ref{fig:interferograms}c). In this simplified model, the population difference corresponds to the photoinduced band occupation variation, which is responsible for triggering the slow insulator-to-metal transformation. In the simulation, the electronic levels are excited by two phase-locked and temporally delayed pulses with Gaussian profiles with input parameters corresponding to the experimental ones. The phenomenological coherence time $T_2$, which represents a dissipative timescale for the off-diagonal coherence ($\rho_{01}$ and $\rho_{10}$ terms), and the bare transition energy $\hbar\omega_{01}$ are left as free effective parameters optimized to match the observed redshift and spectral width variation. The best agreement is obtained for $T_2$=(5.6$\pm$1.0) fs and $\hbar \omega_{01}$=(2.32 $\pm$ 0.02) eV (see Sec. S6 in the SI for more details). {In Sec. S2C we also included in the model the possible change of optical properties induced by the first pump and experienced by the second one. Also in this case a non zero decoherence time is required in order to account for the observed phenomenology.}

The solutions of the OBE also suggest that, for non zero $T_2$, the observed red shift $\Delta$ linearly depends on the detuning $\delta \omega$ between the driving field frequency $\hbar\omega_{p}$ and the transition energy $\hbar\omega_{01}$. Additional experiments, performed at $T$=100~K and varying the central frequency of the pump pulses, confirm the linear relation between $\Delta$ and $\delta \omega$ (Fig. \ref{fig:interferograms}{c} inset). By fitting a linear regression to the experimental values of $\Delta$, we estimate the position of the transition at $\hbar\omega_{01}=\left( 2.37 \pm 0.01 \right)$ eV at $T=\SI{100}{K}$, in good agreement with the value obtained from the OBE numerical results. Taken together, these data demonstrate that the photoinduced IMT after 100 ps is controlled by the initial band population imbalance, which is in turn determined by the coherent dynamics of the many-body state $\left|\Psi \rangle\right.$. The observed $T_2$ is non zero and allows for the exploitation of ultrashort pulses for coherent manipulation of the IMT.

\begin{figure}[t]
\includegraphics[keepaspectratio,clip,width=0.42\textwidth]{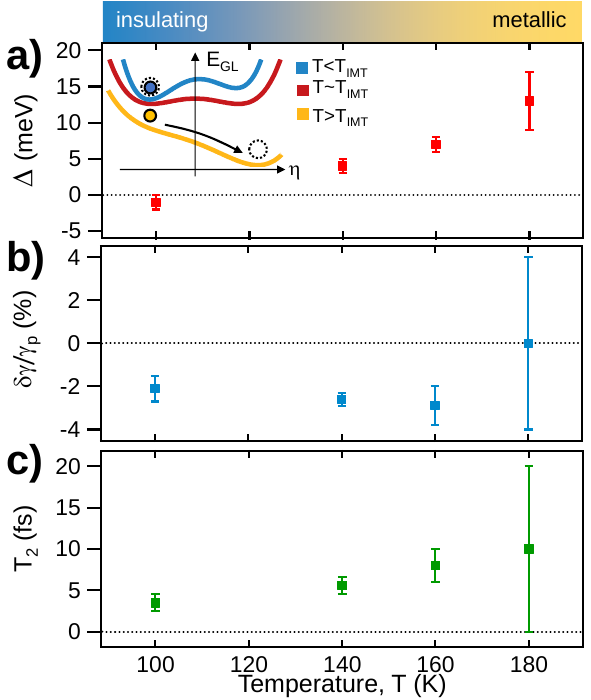}%
\caption{\label{fig:temperature_dependence} {\bfseries Temperature dependence of electronic coherence parameters.} a) Measured energy shift and b) relative spectral width variation as a function of temperature. Inset: Free-energy diagram at different temperature values near the critical temperature $T_{IMT}$. c) Coherence time values as a function of the temperature retrieved by applying the OBE analysis to the experimental results shown in a) and b). Where not reported, the error bar is smaller than the marker size. }
\end{figure}

A natural question arising from the reported evidence of coherent control of the IMT in V$_2$O$_3$ is whether $T_2$ can be increased in order to make coherent manipulation schemes more effective. An intriguing possibility is to exploit the coherent dynamics across the thermally driven transition at $T_{IMT}$. {The flattening of the free-energy curve (see Eq. \ref{eq_freeenergy}) in the vicinity of the spinodal points, where one minimum becomes an inflection point of $E_{GL}$ for $T>T_{IMT}$ (see inset in Fig. \ref{fig:temperature_dependence}a) can give rise to a slowing down of the $\eta$ fluctuations, similarly to second-order phase transitions. Thus, the instantaneous free-energy configuration corresponding to the pump-induced $n_{a_{1g}}$ increase can contribute to enhancing the coherence time that regulates the time evolution of $\left|\Psi \rangle\right.$.}
In Fig. \ref{fig:temperature_dependence}a and b, we report the $\Delta$ and $\delta \gamma/\gamma_p$ values measured for temperatures spanning the 100-180 K range. We note an increasing frequency shift $\Delta$ as $T_{IMT}$ is approached. However, this effect may simply result from the continuous shift of the transition frequency $\hbar \omega_{01}$, which is also suggested by equilibrium optical properties \cite{Qazilbash2008}. In order to retrieve the values of $T_2$, we performed the same analysis, based on OBE, as previously described. The results reported in Fig. \ref{fig:temperature_dependence}c show a moderate increase of the coherence time up to a maximum value $T_2$=(8$\pm$2)~fs. We note that the vanishing of the $\delta R/R (t_1)$ signal at $T\simeq T_{IMT}$ contributes to the large uncertainty of the $T_2$ in the vicinity of the thermally driven IMT.

In conclusion, we have demonstrated the possibility to control the insulator-to-metal transition in the prototypical Mott insulator V$_2$O$_3$ via the coherent manipulation of the $a_{1g}$ and $e^{\pi}_g$ electronic band occupation. The extreme phase stability of the two coherent pump pulses unveils signatures of coherent dynamics, such as detuning and time-broadening of the $\delta R/R$ signal, even for electronic coherence times as short as 5 fs. The electronic coherence shows a tendency to increase in the vicinity of the thermal IMT. These findings suggest that the control of fluctuations of the electronic and structural degrees of freedom is key to enhancing electronic coherence and unlocks the gate for future advances in the coherent manipulation of solid-solid transitions. Experimentally, ultrafast coherent experiments probing the different degrees of freedom (e.g. phonons, spin excitations, orbital fluctuations) are needed to identify the main channels that control the electronic decoherence. This knowledge would indicate the most promising strategies, e.g. different excitation protocols or combination with different external control parameters (chemical composition, strain, pressure, electric/magnetic fields), to preserve the electronic coherence on longer timescales. From a theoretical viewpoint, microscopic models beyond master equations are expected to shed light on the way the many-body systems evolve from a coherent polarization state to a metastable macroscopic phase that can be thermodynamically described by a proper free-energy functional.

\begin{acknowledgments}
C.G., P.F., A.R., A.M., S.M., and S.D.C.  acknowledge financial support from MIUR through the PRIN 2015 (Prot. 2015C5SEJJ001) and PRIN 2017 (Prot. 20172H2SC4\_005) programs. C.G., S.P., and G.F. acknowledge support from Università Cattolica del Sacro Cuore through D.1, D.2.2, and D.3.1 grants. S.M. acknowledges partial financial support through the grant "Finanziamenti ponte per bandi esterni" from Università Cattolica del Sacro Cuore. M.M. acknowledges support from "Severo Ochoa" Programme for Centres of Excellence in R\&D (MINCINN, Grant SEV-2016-0686). M.F. has received funding from the European Research Council (ERC) under the European Union’s Horizon 2020 research and innovation programme, Grant agreement No. 692670 "FIRSTORM". V.P., S.D.C., and G.C. acknowledge support by the European Union Horizon 2020 Programme under Grant Agreement 881603 Graphene Core 3. S.D.C. acknowledges financial support from MIUR through the PRIN 2017 Programme (Prot. 20172H2SC4)
\end{acknowledgments}

\bibliographystyle{unsrt}
\bibliography{bibliography14}

\newpage



\section*{\huge{SUPPLEMENTARY INFORMATION}}

\setcounter{equation}{0}

\setcounter{figure}{0}
\setcounter{section}{0}
\renewcommand{\thefigure}{S\arabic{figure}}
\renewcommand{\thetable}{S\arabic{table}}
\renewcommand{\thesection}{S\arabic{section}}

\section{Supplementary Figure: V$_2$O$_3$ Optical Conductivity}

\begin{figure}[h]
\includegraphics[keepaspectratio,clip,width=0.5\textwidth]{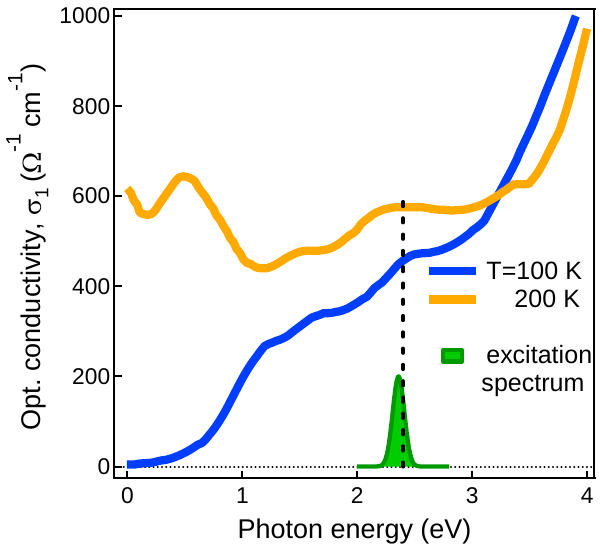}%
\caption{\label{fig:SI_V2O3_optConductivity} {\bfseries Optical Conductivity.} Optical conductivity ($\sigma_1$, real part) as a function of photon energy for the low-temperature ($T=\SI{100}{K}$) antiferromagnetic insulating (AFI, blue solid line) and high-temperature ($T=\SI{200}{K}$) paramagnetic metallic phase (PM, yellow solid line) of a V$_2$O$_3$ thin film sample (the data are taken from Ref. \cite{Qazilbash2008}). The black dashed line highlights the energy position of the interband electronic transition involving the population of the $t_{2g}$ orbitals (at $\sim \SI{2.4}{eV}$) coherently excited by the two phase-locked pump pulses, whose spectrum is schematically reported as green area.}
\end{figure}

\newpage

\section{Optical Bloch Equations Model \label{apx:simulations}}
\subsection{Derivation \label{apx:simulations_derivation}}
The light-matter interaction between a two-level system ($\left|0\rangle\right.$, ground state, and $\left|1\rangle\right.$, excited state) and an external electric field $E \left( t \right)$ is described in terms of the time evolution of the density matix 
$$
\hat{\rho}=\left[ 
\begin{array}{cc}
\rho_{00} & \rho_{01} \\ \rho_{10} & \rho_{11}  
\end{array}
\right],
$$
according to the total Hamiltonian of the two-level system:
\begin{eqnarray} \label{apx:simulations_model_start}
\frac{\partial \hat{\rho}}{\partial t}=\frac{i}{\hbar} \left[ \hat{\rho}, {\mathcal{H}}_{0}+{\mathcal{H}}' \right],
\end{eqnarray}
where 
\begin{eqnarray}
{\mathcal{H}}_{0}=
\left[ 
\begin{array}{cc}
H_{00} & 0 \\ 0 & H_{11}  
\end{array}
\right]
\quad \mbox{and} \quad
{\mathcal{H}}'=
\left[ 
\begin{array}{cc}
0 & -\mu \, E\! \left( t\right) \\ -\mu \, E\! \left( t\right) & 0  
\end{array}
\right]
\end{eqnarray}
are the \emph{unperturbed} and the \emph{interaction} Hamiltonians, respectively, and $\mu$ is the component of the dipole operator parallel to the direction of the electric field $E$.

Including the loss of phase coherence, or \emph{dephasing}, (due to the coupling between the electrons and the environment) and the population relaxation terms, Eq.\,\ref{apx:simulations_model_start} yields to the following set of coupled differential equations (Optical Bloch Equations) \cite{Yariv1989}:
\begin{eqnarray}\label{sist_partenza}
\left\{
\begin{aligned}
\frac{\mbox{d}\rho_{10}}{\mbox{d}t}&=-i \omega_{01} \,  \rho_{10}+\frac{i \mu}{\hbar} \cdot {\Delta \rho} \cdot E\!\left( t \right)-\frac{\rho_{10}}{T_2} \\
\frac{\mbox{d}}{\mbox{d}t} \, {\Delta \rho}&=\frac{2 i \mu \, E\!\left( t \right)}{\hbar} \left( \rho_{10}-\rho_{10}^{*} \right)-\frac{{\Delta \rho}-{{\Delta \rho}_{0}}}{T_1} \\
\end{aligned}
\right.,
\end{eqnarray}
where ${\Delta \rho}=\rho_{00}-\rho_{11}$ is the population difference between the two levels (with $\rho_{ii}$ being the occupation probability of the $i$-th state), $\hbar \omega_{01}=H_{11}-H_{00}$ is the energy difference between the two levels (transition energy), $i$ is the imaginary unit, $T_2$ is the coherence time, ${{\Delta \rho}_{0}}$ is the equilibrium population difference, and $T_1$ is the population decay time.

In our case, the electric field of the two laser pulses is given by 
\begin{eqnarray} \label{eqn:apx_sim_efield1}
E\!\left(t\right) \rightarrow E\!\left(t;t_1\right)=\mathfrak{E}\!\left(t\right)+\mathfrak{E}\!\left(t-t_1\right), \quad \mbox{with} \quad 
\mathfrak{E}\!\left(t\right)=\mathcal{E}\!\left(t\right) \cdot \cos\!\left( {{\omega}}_p \, t \right), 
\end{eqnarray}
with Gaussian envelopes
\begin{eqnarray} \label{eqn:apx_sim_efield2}
\mathcal{E}\!\left(t\right)={\mathscr{E}}_0 \cdot \sqrt[4]{\frac{4 \ln 2}{\pi \tau_{p}^{2}}} \cdot \mathcal{G}\! \left( t\right), 
\quad \mbox{where} \quad
\mathcal{G}\! \left( t\right)= \exp\left[ -\frac{2 \ln 2}{\tau_{p}^{2}} \, t^2 \right].
\end{eqnarray}
In expressions (\ref{eqn:apx_sim_efield1}) and (\ref{eqn:apx_sim_efield2}), ${{\omega}}_p$ is the pump carrier frequency, $t_1$ is the delay between the two pump pulses, ${\mathscr{E}}_0 \cdot \sqrt[4]{\frac{4 \ln 2}{\pi \tau_{p}^{2}}}$ is the amplitude factor (maximum value of the electric field), and $\tau_p$ is the time-duration of the pulse (full width at half maximum of the intensity envelope, $I\left( t\right)$). 

By assuming to be \emph{near resonance} (${{\omega}}_p \sim \omega_{01}$), we introduce the slowly varying variable $\sigma_{10}=\rho_{10} \, e^{i{{\omega}}_p t}$ and we neglect the terms oscillating at $2 {{\omega}}_p$ (rotating wave approximation); finally, we obtain the following set of coupled differential equations:
\begin{eqnarray} \label{eqn:apx_OBE_final_form}
\left\{
\begin{aligned} 
\frac{\mbox{d} \xi}{\mbox{d}t} &=- \left( {{\omega}}_p - \omega_{01} \right) \, \zeta -\Omega_R \, {\Delta \rho}  \, \mathcal{G}\!\left( t-t_1 \right) \; \sin \! \left( {{\omega}}_p \,  t_1 \right)-\frac{\xi}{T_2}\\
\frac{\mbox{d} \zeta}{\mbox{d}t} &=\left( {{\omega}}_p - \omega_{01} \right) \, \xi +\Omega_R \, {\Delta \rho} \, \Big[\mathcal{G}\!\left( t \right)+  \mathcal{G}\!\left( t-t_1 \right) \; \cos \! \left( {{\omega}}_p \, t_1 \right) \Big] -\frac{\zeta}{T_2}\\
\frac{\mbox{d} }{\mbox{d}t} \, {\Delta \rho} &=-4 \Omega_R  \, \Big\{ \zeta \, \mathcal{G}\!\left( t\right) +\Big[ \zeta \cos\!\left( {{\omega}}_p \, t_1 \right)- \xi \sin\!\left( {{\omega}}_p \, t_1 \right)\Big] \,\mathcal{G}\!\left( t-t_1 \right)  \Big\} -\frac{{\Delta \rho}-{{\Delta \rho}_{0}}}{T_1}\\
\end{aligned}
\right.,
\end{eqnarray}
where, $\xi=\mbox{Re}\left\{\sigma_{10}\right\}$, $\zeta=\mbox{Img}\left\{\sigma_{10}\right\}$, and the \emph{Rabi frequency} is defined as
\begin{eqnarray} \label{eqn:apx_Rabi_definition}
\Omega_R=\frac{\mu {\mathscr{E}}_0}{2 \hbar} \, \sqrt[4]{\frac{4 \ln 2}{\pi \tau_{p}^{2}}}.
\end{eqnarray}

The value of ${\mathscr{E}}_0$ is calculated from the fluence according to the following procedure. First, the intensity of a single pulse $I \! \left( t \right)$ is given by \cite{Diels2006}:
\begin{eqnarray}
I\! \left( t \right)=\varepsilon_0 \, c \, n \, \frac{\omega_p}{2 \pi} \int_{t-\frac{\pi}{{{\omega}}_p}}^{t+\frac{\pi}{{{\omega}}_p}} {\mathfrak{E}}^2\!\left(t'\right) \, \mbox{d}t'=\frac{\varepsilon_0 \, c \, n}{2} \, {\mathcal{E}}^2 \left( t \right), 
\end{eqnarray}
where $\varepsilon_0$ is the dielectric constant, $c$ is the speed of light in vacuum, and $n$ is the refractive index (air $\sim 1$). Therefore, in the case of a gaussian field envelope, we obtain
\begin{eqnarray}
I \!\left( t \right)=\frac{\varepsilon_0 \, c \, {\mathscr{E}}_0^2}{2}  \, \sqrt{\frac{4 \ln 2}{\pi \tau_{p}^{2}}} \, \exp\left[ -\frac{4 \ln 2}{\tau_{p}^{2}} \, t^2 \right].
\end{eqnarray}
Finally, the fluence (energy density per unit area) is given by
\begin{eqnarray}
F=\int_{-\infty}^{+\infty} I\!\left( t \right) \, \mbox{d}t=\frac{\varepsilon_0 \, c \, {\mathscr{E}}_0^2}{2}
\end{eqnarray}
in the case of the single pulse. However, in the case the electric field is given by the sum of two delayed pulses, the maximum fluence ($F_0$) is reached when $t_1=0$. Therefore, $F_0=2 \, \varepsilon_0 \, c \, {\mathscr{E}}_0^2$, from which
\begin{eqnarray}
{\mathscr{E}}_0 \left[ \SI{}{V} \cdot \SI{}{s^{1/2}} \cdot \SI{}{m^{-1}} \right]=\SI{13.73}{} \cdot \sqrt{F_0 \left[ \SI{}{ m^{-2}} \cdot   \SI{}{ m^{-2}} \right]}
\end{eqnarray}

In our condition, $2 \pi / \Omega_R \gg \tau_p$. Moreover, we consider a linear regime, in which the results are independent from the incident fluence.

\subsection{Numerical Solution of Optical Bloch Equations}

\begin{figure}[h]
\centering
\includegraphics[keepaspectratio,clip,width=\textwidth]{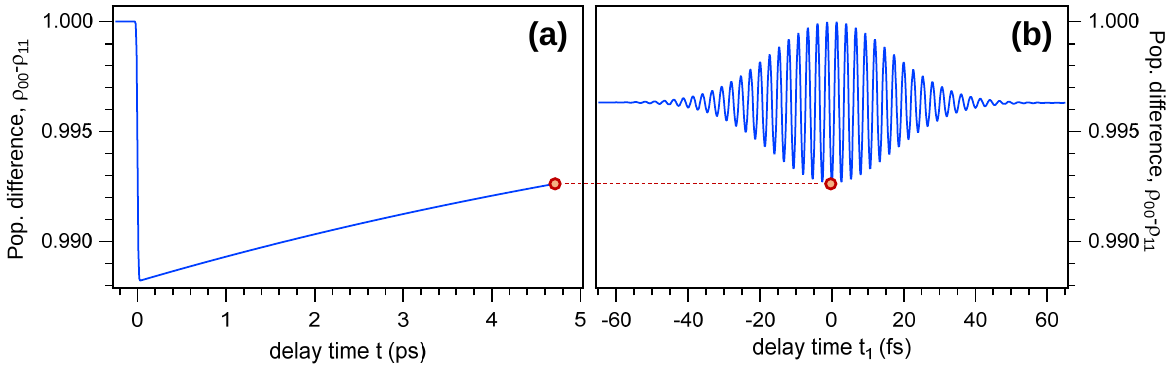}
\caption{\textbf{Time-Dependent Population Dynamics.} (a) Temporal evolution of the population difference excited by two phase-coherent pump pulses at fixed $t_1$ value. (b) Population difference as a function of the time delay between the pump pulses ($t_1$) at large delay time values ($t_{2} \sim \SI{5}{ps}$). The numerical results have been obtained by assuming a transform-limited pulse duration $\tau_p=\SI{20}{fs}$ at $\hbar {{\omega}}_p=\SI{1.550}{eV}$, a coherence time value $T_2=\SI{0.1}{fs}$, and a transition energy $\hbar \omega_{01}=\SI{1.597}{eV}$. In panel a, we assumed $t_1=0$. The red marker highlights the population difference value at large $t_2$ values when the two-level system is excited by two phase-coherent pump pules arriving at the same time ($t_1=0$). 
}
\label{fig:ITRPP_simulation_popDynamics}
\end{figure}

\begin{figure}[t]
\centering
\includegraphics[keepaspectratio,clip,width=\textwidth]{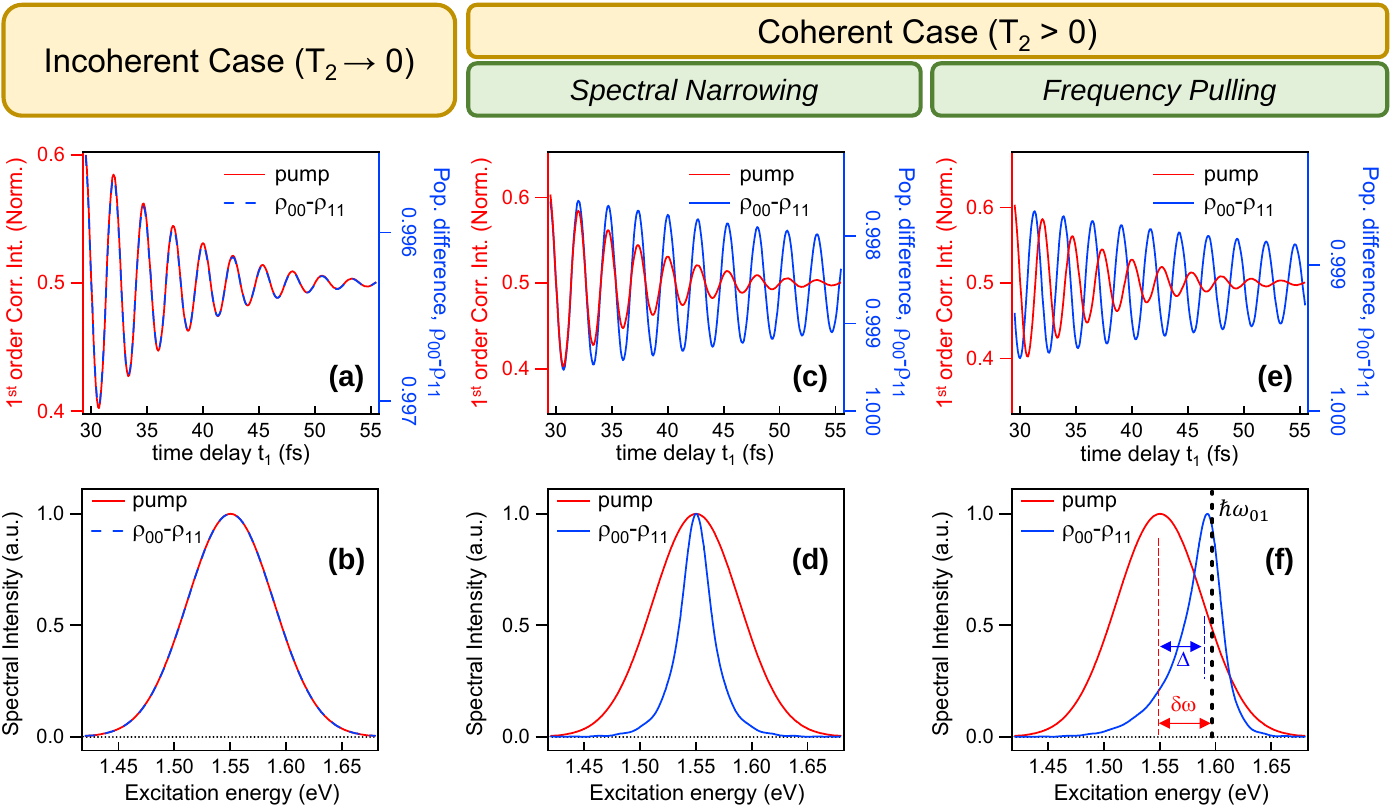}
\caption{\textbf{Spectral Narrowing and Frequency Pulling.} Temporal (panels a, c, and e) and spectral (panels b, d, and f) profiles of the first-order correlation function of the excitation (red line) and the population difference (blue line) for the incoherent (a and b) and coherent (c, d, e, and f) superposition of $\left| 0\rangle \right.$ and $\left| 1\rangle \right.$ states of the two-level system excited by two phase-coherent pulses. The numerical results have been obtained by assuming a transform-limited pulse duration $\tau_p=\SI{20}{fs}$ at $\hbar {{\omega}}_p=\SI{1.550}{eV}$ (red dashed line in panel f). In the coherent case, $T_2=\SI{40}{fs}$. In panels e and f, the transition energy was set to $\hbar \omega_{01}=\SI{1.597}{eV}$ (black dotted line in panel f), corresponding to a detuning $\delta \omega=\SI{-47}{meV}$. $\Delta$ is the shift ($\Delta \simeq \SI{-39}{meV}$ in panel f). In panels b, d, and f, the horizontal axis is the excitation energy: the transform variable of the time delay $t_1$.
}
\label{fig:ITRPP_simulation_over}
\end{figure}

\begin{figure}[t]
\centering
\includegraphics[keepaspectratio,clip,width=\textwidth]{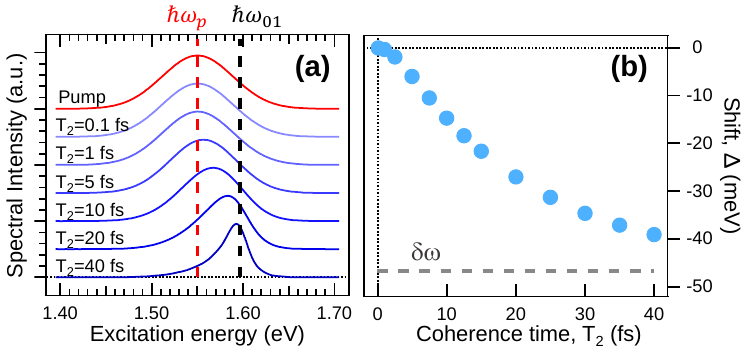}
\caption{\textbf{Coherence Time dependent Shift.} Dependence of $\Delta$ on $T_2$ at fixed $\delta \omega$ value. (a) Pump spectrum (red dashed line) compared to signal spectrum (blue solid line) obtained for different $T_2$ values (see labels). (b) $\Delta$ vs $T_2$ curve retrieved from the spectra in panel a. Here, the data are obtained from the numerical solution of OBE with $\tau_p=\SI{20}{fs}$, $\hbar {{\omega}}_p=\SI{1.550}{eV}$ (red dashed line in panel a), and $\hbar \omega_{01}=\SI{1.597}{eV}$ (black dashed line in panel a); therefore, $\delta \omega=-\SI{47}{meV}$ (gray dashed line in panel b).
}
\label{fig:ITRPP_simulation_shift_vs_cohTime}
\end{figure}

In this section, we discuss the effect of the coherence on the dynamics of the population difference (${\Delta \rho}=\rho_{00}-\rho_{11}$) of a two-level system excited by two phase-locked pump pulses. The results reported in Figs. \ref{fig:ITRPP_simulation_popDynamics}-\ref{fig:ITRPP_simulation_shift_vs_cohTime} have been obtained by numerically solving the Optical Bloch Equations (OBE) in Eq.\,\ref{eqn:apx_OBE_final_form} for a transform-limited pulse whose duration (FWHM of the temporal intensity profile) is $\tau_p=\SI{20}{fs}$. 

In order to obtain the population difference as a function of the delay time $t_1$ at large $t_2$ values, we adopted the following procedure. First, we solve the OBE system in Eq.\,\ref{eqn:apx_OBE_final_form} at fixed $t_1$ value. This provides $\left(\rho_{00}-\rho_{11}\right)(t)$, which is the population difference as a function of the time delay $t$ (we set $t=0$ when the first pump pulse excites the two-level system, as shown in Fig. 1), as depicted in Fig. \ref{fig:ITRPP_simulation_popDynamics}a in the case $t_1=0$. We extracted the population difference value at large delay time $t$ (red marker in Fig. \ref{fig:ITRPP_simulation_popDynamics} at $t\sim \SI{5}{ps}$), which corresponds to the population difference value at a large delay time $t_2$; indeed, we assumed $t_2=t-t_1$. Then, the same procedure is repeated for a different $t_1$ values; therefore, we obtain $\left(\rho_{00}-\rho_{11}\right)(t_1)$, as shown in Fig. \ref{fig:ITRPP_simulation_popDynamics}b.

The coherence effects on the two-level system population are described in Fig. \ref{fig:ITRPP_simulation_over}, where we compare the $t_1$-dependent population difference profile and the first-order correlation function (interferogram) of the pump pulses. 

First, we consider the \emph{incoherent} case (panels \ref{fig:ITRPP_simulation_over}a and b) corresponding to the condition in which the coherence between the $\left| 0\rangle \right.$ and $\left| 1\rangle \right.$ states decays instantaneously ($T_2 \rightarrow 0$). Here, the population difference (blue dashed curve in panel \ref{fig:ITRPP_simulation_over}a) evolves with the same oscillation frequency and the same temporal envelope as the first-order correlation function of the pump excitation (red solid curve). This behaviour is confirmed by taking the Fourier-Transform (FT) of the two interferograms: the resulting spectra perfectly overlap (panel \ref{fig:ITRPP_simulation_over}b). 

The situation is different in the \emph{coherent} case, where the coherent superposition of $\left| 0\rangle \right.$ and $\left| 1\rangle \right.$ states is maintaned on a non-vanishing timescale ($T_2>0$). As shown in panel \ref{fig:ITRPP_simulation_over}c, although the two interferograms oscillate in phase, the $\rho_{00}-\rho_{11}$ interferogram exhibits a temporal envelope that is significantly longer than the pump interferogram. This width difference is confirmed also by the FT of the two waveforms, displayed in panel \ref{fig:ITRPP_simulation_over}d. Indeed, the FT of $\rho_{00}-\rho_{11}$ (which will be labelled as \emph{signal} spectrum \footnote{This because the experimental signal (relative reflectivity variation $\delta R/R$ in our experiment) is proportional to $\rho_{00}-\rho_{11}$}) is narrower than the pump spectrum. This property, which we label as \emph{spectral narrowing}, originates from the non-zero lifetime of the quantum coherence between the $\left| 0\rangle \right.$ and $\left| 1\rangle \right.$ states. 

The case described in Fig. \ref{fig:ITRPP_simulation_over}c and d corresponds to the situation in which the carrier frequency of the external excitation matches the frequency of the transition: ${{\omega}}_p \equiv \omega_{01}$ (\emph{resonant} case). Here, in addition to spectral narrowing, a second effect, labelled \emph{Frequency Pulling}, may occur when the external field is non-resonant to the transition, $\delta \omega$=$  \hbar {{\omega}}_p-\hbar \omega_{01}$ being the detuning, and provided $T_2$ being finite. Fig. \ref{fig:ITRPP_simulation_over}e shows the case of transition energy higher than the excitation energy (negative detuning, $\delta \omega<0$). We observe that the $\rho_{00}-\rho_{11}$ interferogram oscillates with a frequency larger than the pump one. This difference in the oscillation frequency is confirmed also by the spectra in Fig. \ref{fig:ITRPP_simulation_over}f. Indeed, the FT of $\rho_{00}-\rho_{11}$ (signal spectrum) is blue-shifted with respect to the pump spectrum thereby approaching the energy position of the transition $\hbar \omega_{01}$ (dashed dotted line in Fig. \ref{fig:ITRPP_simulation_over}f).

To quantify the coherence effect, we introduce the energy shift $\Delta$ (see Fig. \ref{fig:ITRPP_simulation_over}f), defined as the spectral distance between the pump spectrum (peaked at $\hbar {{\omega}}_p$) and the signal spectrum (peaked at $\bar{{E}}^{}_{\, \scriptsize{\mbox{sig}}}$): $\Delta=\hbar {{\omega}}_p-\bar{{E}}^{}_{\, \scriptsize{\mbox{sig}}}$.

To clarify the relation between $\Delta$ and $\delta \omega$, we analyzed the dependence of $\Delta$ on $T_2$ at fixed $\delta \omega$ value and the results are displayed in Fig. \ref{fig:ITRPP_simulation_shift_vs_cohTime}. In particular, panel \ref{fig:ITRPP_simulation_shift_vs_cohTime}a compares the pump spectrum (red dashed line) and the signal spectra (blue continuous lines) obtained for different $T_2$ values. For each $T_2$ value, the $\Delta$ value is extracted. The dependence of $\Delta$ on $T_2$ is summarized in panel \ref{fig:ITRPP_simulation_shift_vs_cohTime}b. The numerical results suggest that $\Delta$ depends on $T_2$ and, in general, $|\Delta| < |\delta \omega|$. Moreover, for large values of $T_2$, $\Delta \approx \delta \omega$.

\newpage
{\subsection{Alternative Model: Time-dependent Dipole Moment}}
{As discussed in Sec. \ref{apx:simulations_derivation}, the theoretical model we adopted to analyze the experimental results takes into account two main contributions: the coherence between the two levels (whose corresponding parameter is the coherence time $T_2$) and the detuning ($\delta \omega$) between the pump and the electronic transition excited. In the present section we describe an alternative approach in order to investigate whether the spectral narrowing experimentally observed could originate from a change in the refractive index induced by the the combined effect of two time-delayed pump pulses, rather than from electronic coherence.}

\begin{figure}[t]
\includegraphics[keepaspectratio,clip,width=\textwidth]{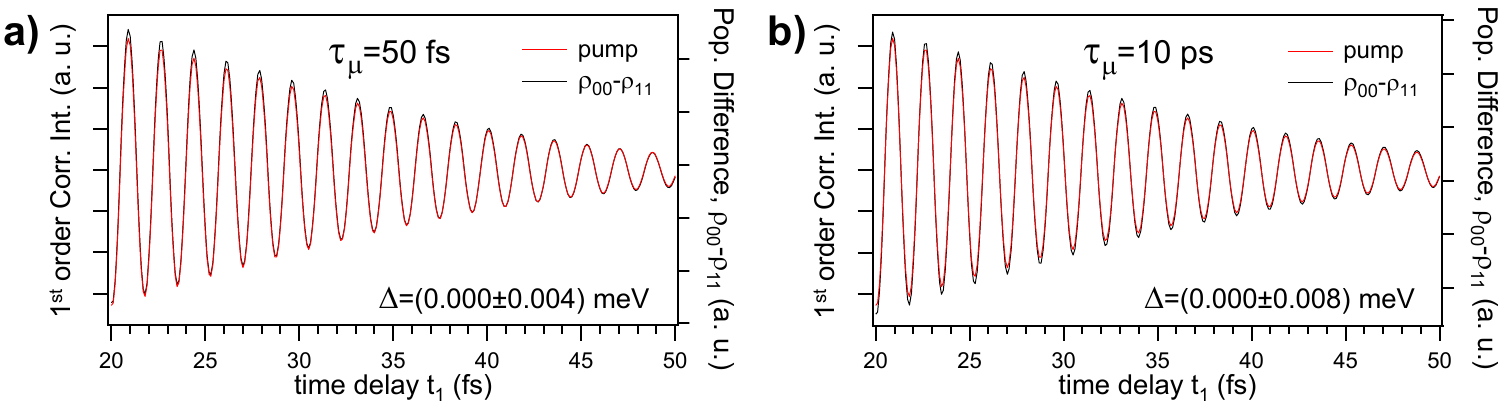}%
\caption{\label{fig:interferogram_TDDM} {{\bfseries Time-dependent Dipole Moment.} (a-b) Calculated two-level population difference as a function of the pump-pump delay $t_1$ (black line) compared to the first-order correlation function of the pump pulse (red line). The decay time of the dipole-moment variation is $\tau_{\mu}=\SI{50}{fs}$ (a) and $\tau_{\mu}=\SI{10}{ps}$ (b). The simulations have been performed by solving OBE in Eq.\,\ref{eqn:sistema_2gaussian_pulses_mom_dipolo_0926} with $T_2 \approx 0$ and $\delta \omega=\SI{0.06}{eV}$.}}
\end{figure}

In the second approach, we keep the hypothesis of a two-level system excited by two coherent pump pulses. However,  we do not include the role of the coherence between the two levels (thus we set $T_2 \rightarrow 0$) and we assume a different dipole moment value for the two pump pulses. This represents the fact that the first pump pulse \emph{sees} the system at equilibrium (ground state), while the second pump pulse \emph{finds} the system in an excited state. Therefore, we replace the interaction Hamiltonian 
\begin{eqnarray}
{\mathcal{H}}'=-\hat{\mu} \cdot E \! \left( t \right)=-\mu \cdot \mathfrak{E}\!\left(t\right)-\mu \cdot \mathfrak{E}\!\left(t-t_1\right).
\end{eqnarray}
with
\begin{eqnarray} \label{eqn:interaction_term}
{\mathcal{H}}'=-\mu_1 \cdot \mathfrak{E}\!\left(t\right)-\mu_2 \cdot \mathfrak{E}\!\left(t-t_1\right),
\end{eqnarray}
where the dipole moment $\mu_1=\mu$ ($\mu_2=\mu+\delta \mu$) describes the interaction between the system and the first (second) pulse. In the most general case, (\emph{i}) we consider a time-dependent dipole moment variation, $\delta \mu=\delta \mu \! \left( t \right)$, and (\emph{ii}) we assume the maximum amplitude of $\delta \mu$ as the difference between the dipole moment value in the metallic ($\mu^{\scriptsize{\mbox{met}}}$) and insulating phase ($\mu$). Therefore, we assume
\begin{eqnarray}
\mu_2=\mu_2 \! \left( t \right)=\mu+\Theta \! \left( t \right) \cdot \left( \mu^{\scriptsize{\mbox{met}}} - \mu \right) \cdot e^{-t/\tau_{\mu}},
\end{eqnarray}
where $\tau_{\mu}$ is the decay time of the dipole moment variation and $\Theta \! \left( t \right)$ is the Heaviside step function. As in the previous model, we assume $\mu \propto \sqrt{\varepsilon_{2}^{\scriptsize{\mbox{ins}}}} \propto \sqrt{\sigma_{1}^{\scriptsize{\mbox{ins}}}}$ \cite{Yariv1989} and, consequently, $\mu^{\scriptsize{\mbox{met}}}\propto \sqrt{\sigma_{1}^{\scriptsize{\mbox{met}}}}$. Within this framework, the interaction term in Eq.\,\ref{eqn:interaction_term} takes the form
\begin{eqnarray}
{\mathcal{H}}'=-\mu \cdot \Big[ \mathfrak{E}\!\left(t\right)+\psi\!\left( t \right) \cdot \mathfrak{E}\!\left(t-t_1\right) \Big],
\end{eqnarray}
where
\begin{eqnarray}
\psi\!\left( t \right)=1+\Theta \! \left( t \right) \cdot {\left( \sqrt{\frac{\sigma_{1}^{\scriptsize{\mbox{met}}}}{\sigma_{1}^{\scriptsize{\mbox{ins}}}}}-1 \right)} \cdot e^{-t/\tau_{\mu}}.
\end{eqnarray}
Therefore, we numerically solved the following set of coupled differential equations
\begin{eqnarray}\label{eqn:sistema_2gaussian_pulses_mom_dipolo_0926}
\left\{
\begin{aligned}
\frac{\mbox{d} \xi}{\mbox{d}t} &=- \left( \omega_p - \omega_{01} \right) \, \zeta -\Omega_R \cdot {\Delta \rho} \cdot \psi \! \left( t \right) \cdot \mathcal{G}\!\left( t-t_1 \right) \cdot \sin \! \left( \omega_p \, t_1 \right)-\frac{\xi}{T_2}\\
\frac{\mbox{d} \zeta}{\mbox{d}t} &=\left( \omega_p - \omega_{01} \right) \, \xi +\Omega_R \, {\Delta \rho} \, \Big[\mathcal{G}\!\left( t \right)+\psi \! \left( t \right) \cdot   \mathcal{G}\!\left( t-t_1 \right) \cdot \cos \! \left( \omega_p \,  t_1 \right) \Big] -\frac{\zeta}{T_2}\\
\frac{\mbox{d} }{\mbox{d}t} \, {\Delta \rho} &=-4 \Omega_R  \, \Big\{ \zeta \, \mathcal{G}\!\left( t\right) +\Big[ \zeta \cos\!\left( \omega_p \, t_1 \right)- \xi \sin\!\left( \omega_p \, t_1 \right)\Big] \cdot \psi \! \left( t \right) \cdot \mathcal{G}\!\left( t-t_D \right)  \Big\} -\frac{{\Delta \rho}-{{\Delta \rho}_{0}}}{T_1}\\
\end{aligned}
\right.
\end{eqnarray}
in the limit $T_2 \rightarrow 0$ and $\delta \omega>0$. The value of $\sigma_1^{\scriptsize{\mbox{ins}}}$ and $\sigma_1^{\scriptsize{\mbox{met}}}$ at $\hbar \omega_p$ are taken from Ref. \cite{Qazilbash2008} (see Fig. \ref{fig:SI_V2O3_optConductivity}). {Fig. \ref{fig:interferogram_TDDM} shows the population difference dynamics obtained for two values of the decay time of the dipole moment variation: $\tau_{\mu}=\SI{50}{fs}$ (a) and $\tau_{\mu}=\SI{10}{ps}$ (b). The reported results demonstrate that the contribution to both $\Delta$ and $\delta \gamma/\gamma_p$ given by a finite value of $\tau_{\mu}$ is negligible.}

{In conclusion, this complementary analysis suggests that the hypothesis of an interaction Hamiltonian as in Eq.\,\ref{eqn:interaction_term} is not capable to reproduce the main experimental results, which are characterized by a simultaneous spectral shift and negative relative width variation (althought it might be adopted for a better estimation of the coherence time and detuning). Therefore, this reinforces the importance of including the coherence between the two levels (\emph{i.e.}, considering the parameter $T_2>0$) in the analysis (as discussed in Sec. \ref{apx:simulations_derivation}). }


\clearpage

\newpage
\section{Choice of the Experimental Parameters to Optimize Coherent Effects}

In this section, we discuss the dependence of the shift ($\Delta$) from two experimental parameters: the detuning ($\delta \omega$) and the pulse bandwidth ($\Delta E$, which direcly derives from the pulse duration due to the transform-limited, TL, condition). This represents an important guideline in the choice of the experimental parameters in order to maximize the effects of coherence in the experiments. The results described in the following have been obtained from the numerical solution of Optical Bloch Equations (see Sec. \ref{apx:simulations} for more details), at fixed coherence time ($T_2$). 

\begin{figure}[t]
\includegraphics[keepaspectratio,clip,width=0.5\textwidth]{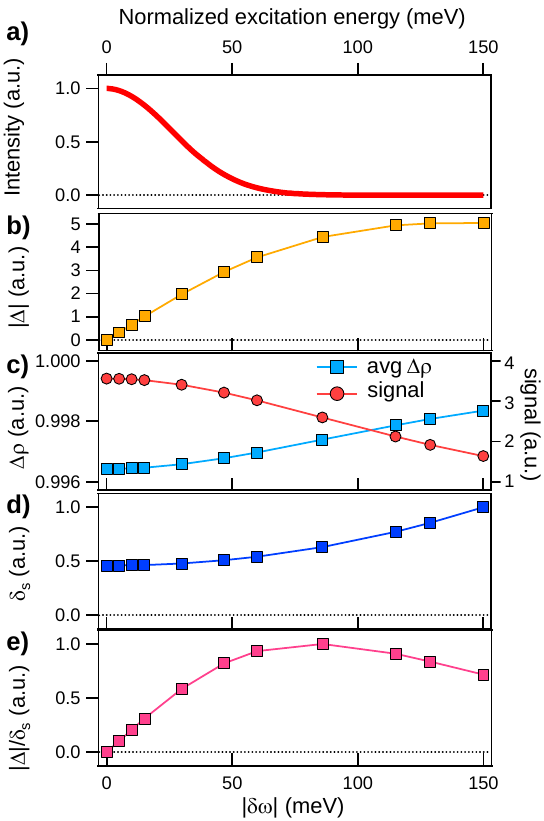}%
\caption{\label{fig:V2O3_ITRPP_introDet} {\bfseries Maximum Shift vs Detuning.} (a) Pump spectral intensity as a function of the normalized excitation energy. Detuning dependence of (b) shift {(absolute value ) $\left|\Delta\right|$}, (c) average population difference ${\Delta \rho}$, (d) signal error $\delta_s$, and (e) ratio {$\left|\Delta\right| / \delta_s$}. These results are obtained from simulations performed at fixed coherence time $T_2=\SI{5}{fs}$, pulse duration $\tau_p=\SI{30}{fs}$ (TL), and maximum fluence $\SI{0.5}{}$ mJ/cm$^2$.}
\end{figure}

First, we focus on the detuning and we consider an excitation pulse whose spectrum is shown in Fig. \ref{fig:V2O3_ITRPP_introDet}a. Since this argument is general, the spectral intensity is reported as a function of the \emph{normalized excitation energy} ($\hbar {{\omega}}-\hbar {{\omega}}_p$) in order to avoid the choice of a specific central frequency $\omega_p$. As reported in Fig. \ref{fig:V2O3_ITRPP_introDet}b, the shift increases with increasing detuning, (in this picture we considered only the cases of transition energy higher than the pump energy). Therefore, one may suggest to adopt a pump with central frequency far from the transition to be excited (large detuning). 
However, it is important to consider the population difference ${\Delta \rho}=\rho_{00}-\rho_{11}$ (blue square markers in \ref{fig:V2O3_ITRPP_introDet}c) and, in particular, the occupation of the two-levels. In our definition, ${\Delta \rho}=1$ corresponds to a population occupying only the ground state $\left| 0 \right.\rangle$ ($\rho_{00}=1$) and this is the value before the arrival of the excitations (negative delay, ${{\Delta \rho}_{eq}}$). Therefore, as shown by \ref{fig:V2O3_ITRPP_introDet}c, this means that the occupation of $\left| 1 \right.\rangle$ and, consequently, the population imbalance (${{\Delta \rho}_{eq}}-{\Delta \rho}$), induced by the pump pulses, decreases when the detuning increases. This behaviour suggests that the output signal of the experiment, which (as a first approximation) is modelled as proportional to ${{\Delta \rho}_{eq}}-{\Delta \rho}$, decreases with increasing detuning (red circular markers in Fig. \ref{fig:V2O3_ITRPP_introDet}c). By assuming that the error on the signal ($\delta_s$) is inversely proportional to the signal itself ($\delta_s \propto 1/({{\Delta \rho}_{eq}}-{\Delta \rho})$), it follows that the error on the signal increases with increasing detuning, as depicted in Fig. \ref{fig:V2O3_ITRPP_introDet}d. Therefore, it follows that the maximum detectable shift is given by a compromise between the detuning and the error on the signal. To estimate this, we calculated the ratio $\Delta/\delta_s$ (Fig. \ref{fig:V2O3_ITRPP_introDet}e). Therefore, in order to maximize the value of the shift detected in the experiment, at fixed $T_2$, the values of the detuning providing the highest $\Delta/\delta_s$ ratio should be chosen.

\begin{figure}[t]
\includegraphics[keepaspectratio,clip,width=0.5\textwidth]{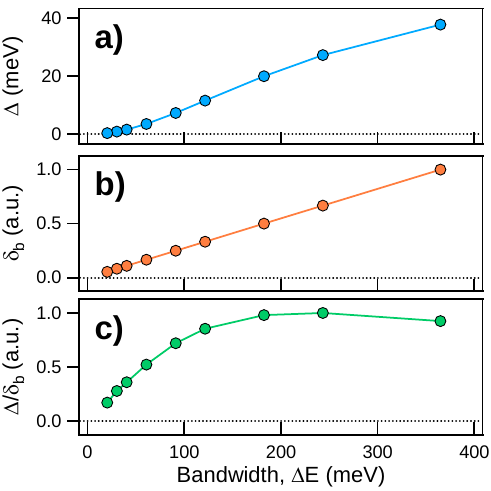}%
\caption{\label{fig:V2O3_ITRPP_introBand} {\bfseries Maximum Shift vs Bandwidth.} Bandwidth dependence of (a) shift $\Delta$, (b) shift error $\delta_b$, and (e) ratio $\Delta / \delta_b$. These results are obtained from simulations performed at fixed coherence time $T_2=\SI{5}{fs}$ and detuning $\delta \omega=\SI{0.06}{eV} $.}
\end{figure}

Second, we consider the role of the pulse bandwidth. As displayed in Fig. \ref{fig:V2O3_ITRPP_introBand}a, at fixed $T_2$, the shift increases with increasing spectral bandwidth, \emph{i.e.} increases with decreasing pulse duration (transform-limited, TL, condition). Therefore, one may suggest to employ a broadband (\emph{i.e} ultrafast) pulse to detect a large shift. Indeed, $\Delta$ vanishes in the case of pulses whose time duration is much longer than the coherence time. However, in the case of an ultrafast pulse, the ultrabroad bandwidth may hinder the correct estimation of the spectrum central position, thus making the calculation of the shift more difficult. In time-domain description, in the case of an ultrafast pulse, the time-domain interferogram displays only few oscillations within the temporal envelope. Althought the pump and signal interferograms may oscillate at different frequencies, the small number of optical cycles prevents the precise evaluation of the shift. In this case, therefore, it is reasonable to assume that the error on the shift ($\delta_b$) is proportional to the pulse bandwidth: $\delta_b \propto \Delta E$, as reported in Fig. \ref{fig:V2O3_ITRPP_introBand}b. Therefore, in order to maximize the value of the shift detected in the experiment, at fixed $T_2$, the values of the bandwidth providing the highest $\Delta/\delta_b$ ratio (displayed in Fig. \ref{fig:V2O3_ITRPP_introBand}c) should be chosen.

In conclusion, based on these considerations, we can state that, for a fixed value of $T_2$, there are optimum values of detuning and bandwidth in order to optimize the coherence effects. According to the results shown in Figs. \ref{fig:V2O3_ITRPP_introDet} and \ref{fig:V2O3_ITRPP_introBand}, the optimum region\footnote{The limits of the optimum region are identified as the points whose vertical value is 90\% of the maximum of the curves.} corresponds to a detuning range of $\SI{55}{}-\SI{115}{meV}$ and a bandwidth range of $\SI{140}{}-\SI{160}{meV}$ ($\SI{5}{}-\SI{13}{fs}$ pulse duration). 

\clearpage

\newpage
\section{Experimental Setup}
\begin{figure}[t]
\includegraphics[keepaspectratio,clip,width=0.8\textwidth]{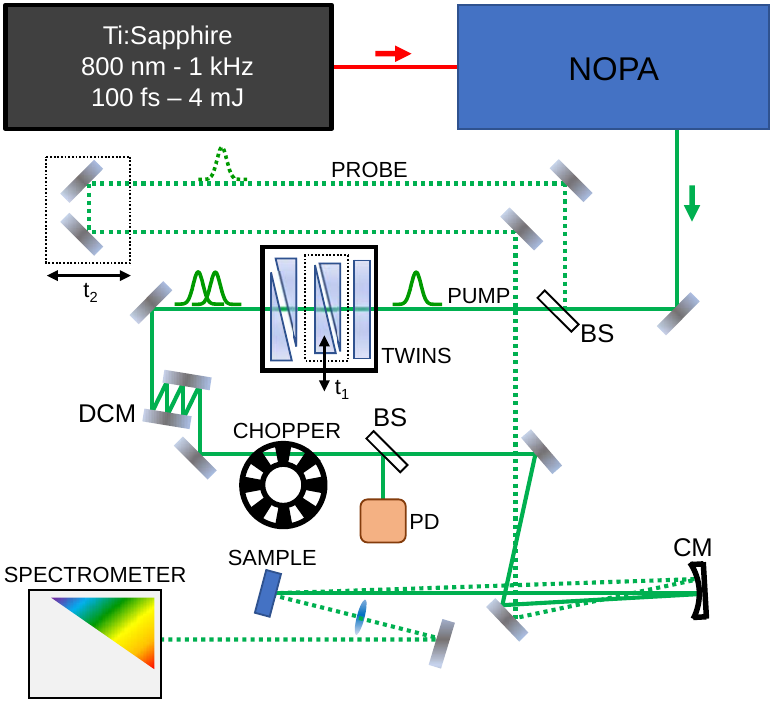}%
\caption{\label{fig:img_SETUP} {\bfseries Setup.} Sketch of the experimental setup: NOPA, Non-collinear Optical Parametric Amplifier; BS, beam splitter; DCM, double chirped mirrors; TWINS, Translating-Wedge-Based Identical Pulses eNcoding System; PD: photodiode; CM, concave mirror.}
\end{figure}

A scheme of experimental setup (consisting in a two-dimensional electronic spectrometer) is shown in Fig. \ref{fig:img_SETUP}. It is based on a Ti:Sapphire laser system which delivers $\SI{100}{fs}$-long IR (around $\SI{800}{nm}$) pulses with $\SI{4}{mJ}$ energy at $\SI{1}{kHz}$ repetition rate. A portion of the laser output is used to pump a Non-collinear Optical Parametric Amplifier (NOPA), which produces broadband pulses in the UV-VIS region. The NOPA output pulses are compressed to nearly TL duration by multiple bounces on double chirped mirrors (DCMs). A beam-splitter divides the NOPA output into two beams: the pump and the probe, whose relative delay $t_2$ is tuned with a motorized linear delay stage. Two phase-coherent and delayed replicas are generated from the pump beam thanks to a common-path birefringent interferometer, called Translating-Wedge-Based Identical Pulses eNcoding System (TWINS) \cite{Brida2012}. The additional dispersion introduced by the device is corrected by an additional pair of DCMs. The relative delay between the two pump replicas, $t_1$, is monitored by a photo-diode (PD), which records the intensity as a function of $t_1$ (pump interferogram). The pump beam is modulated by a mechanical chopper locked at $\SI{500}{Hz}$. The pump and the probe beams are focused onto the sample in a non-collinear beam geometry by a concave mirror (CM). The spatial overlap and the spot sizes are monitored by a CCD camera (typical dimensions, fwhm: $\SI{60}{}$ $\mu\SI{}{m}$ and $\SI{120}{}$ $\mu\SI{}{m}$ for probe and pump, respectively). After the interaction with the sample, the broadband reflected probe is detected by a spectrometer, thus enabling the resolution with respect to the detection energy ($\hbar \omega_3$). At fixed delays ${\bar{t}}_1$ and ${\bar{t}}_2$, the frequency-resolved relative reflectivity variation $\delta R/R \! \left( \hbar \omega_3; {\bar{t}}_1, {\bar{t}}_2  \right)$ is obtained by subtracting two consecutive spectra, the first recorded with the pump on ($R_{\scriptsize{\mbox{on}}}$) and the second with the pump blocked ($R_{\scriptsize{\mbox{off}}}$), and then dividing by the equilibrium reflectivity: $$\frac{\delta R}{R} \! \left( \hbar \omega_3; {\bar{t}}_1, {\bar{t}}_2  \right)=\frac{R_{\scriptsize{\mbox{on}}}\!\left( \hbar \omega_3; {\bar{t}}_1, {\bar{t}}_2 \right)-R_{\scriptsize{\mbox{off}}}\!\left( \hbar \omega_3; {\bar{t}}_1, {\bar{t}}_2  \right)}{R_{\scriptsize{\mbox{off}}}\!\left( \hbar \omega_3; {\bar{t}}_1, {\bar{t}}_2  \right)}.$$ If this procedure is repeated for each time delay $t_1$, the resolution of the signal with respect to the excitation energy is obtained by computing the Fourier-Transform with respect to $t_1$. The signal dynamics is obtained by repeating this procedure for various $t_2$ values.

The shift resolution achievable in our experiment is estimated from the noise at fixed $t_1$-delay. In our setup, the sources of noise are the laser intensity and phase stability fluctuations, and they are compatible with a shift resolution of $\SI{2}{meV}$.

\clearpage
\newpage
\section{Analysis of the Measurements at Different Temperatures}

\begin{figure}
\centering
\includegraphics[keepaspectratio,clip,width=0.8\textwidth]{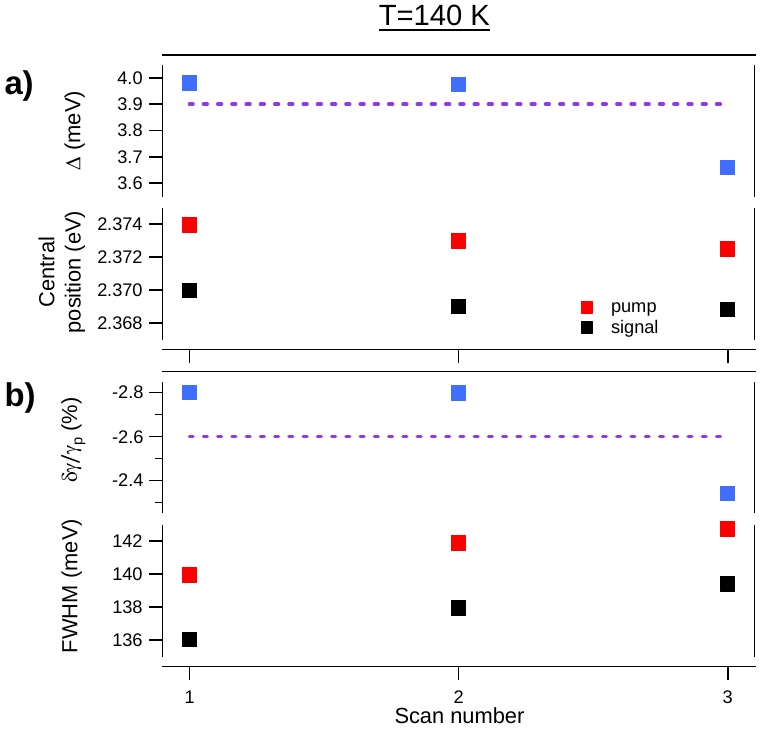}
\caption{\textbf{Statistical Analysis of the Measurement at $\mathsf{T}=\SI{140}{K}$.} (a) Shift (blue markers), pump and signal spectrum central position (red and black markers, respectively) as function of the scan. (b) Relative width variation (blue markers), pump and signal spectrum width (red and black markers, respectively) as function of the scan number. The violet dashed line highlights the average value.}
\label{fig:apx_V2O3_ITRPP_140K_statAnalysis}
\end{figure}

In order to retrieve the shift ($\Delta$) and the relative width variation ($\delta \gamma / \gamma_p$), the following procedure was adopted:
\begin{itemize}
    \item The measurement taken at fixed temperature is obtained by averaging consecutive acquisitions (each of those is labelled as \emph{scan}).
    \item The number of scans (at fixed temperature) was chosen in order to obtain an error on the shift of the order of $\SI{1}{meV}$, except in the case of measurement taken at $T=\SI{180}{K}$ (for which it is not possible to go below $\SI{4}{meV}$ due to a much lower signal-to-noise ratio in the paramagnetic metallic phase of vanadium sesquioxide compared to the antiferromagnetic insulating one).
    \item In each scan, the output data consists in two interferograms: the first-order correlation function of the pump excitation (\emph{pump} interferogram) and the $\delta R/R$ signal (\emph{signal} interferogram).
    \item The interferograms were Fourier-Transformed with respect to $t_1$ to obtained the corresponding pump ($p$) spectrum and signal ($s$) spectrum. 
    \item For both pump and signal spectra, we calculated the central value position ($\hbar {{\omega}}_p$ and $\bar{{E}}^{}_{\, \scriptsize{\mbox{sig}}}$, respectively) and the width ${\gamma}_i$, where $i={p}$ or ${s}$, given by the full-width at half maximum. 
    \item For each scan, the values of the shift and the relative width variation are obtained as
    $$
    \Delta=\hbar {{\omega}}_p-\bar{{E}}^{}_{\, \scriptsize{\mbox{sig}}}
    \qquad
    \mbox{and}
    \qquad
    \delta \gamma/\gamma_p=\frac{{\gamma}_s}{{\gamma}_p}-1,
    $$
    respectively. As an example, see Fig. \ref{fig:apx_V2O3_ITRPP_140K_statAnalysis}.
    \item The average values of the shift and the relative width variation were finally obtained. As an example, see the dashed violet line in panels a and b in Fig. \ref{fig:apx_V2O3_ITRPP_140K_statAnalysis}.
\end{itemize}


\clearpage
\newpage

\newpage
\section{Temperature-dependent Coherence Time.}
\label{apx:subsec_V2O3_ITRPP_2p4eV_procedure_T2}

\begin{figure}[h]
\centering
\includegraphics[keepaspectratio,clip,width=0.9\textwidth]{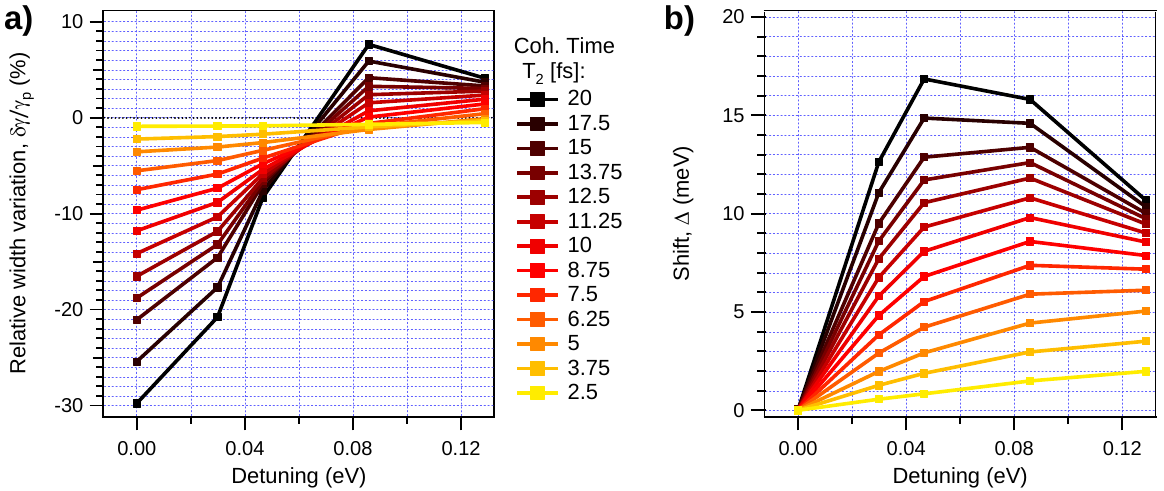}
\caption{\textbf{Simulations.} (a) Relative width variation, $\delta \gamma/\gamma_p$, and (b) normalized shift, $\Delta_N$, as a function of the detuning, for various $T_2$ values. The data are obtained from the numerical solution of OBE.}
\label{fig:apx_V2O3_ITRPP_2p4eV_procedure_T2}
\end{figure}

In this section, we describe the procedure adopted to extract the temperature dependence of the coherence time $T_2$ ({Fig. 3c}) starting from the energy shift $\Delta$ and relative width variation $\delta \gamma/\gamma_p$ obtained experimentally ({Fig. 3a and b}, respectively):

\begin{itemize}
    \item Within the framework of the OBE simulations, we numerically calculated the calibration curves $\Delta$ vs detuning and $\delta \gamma/\gamma_p$ vs detuning, for different values of the coherence time $T_2$. As an example, the calibration curves are displayed in Fig. \ref{fig:apx_V2O3_ITRPP_2p4eV_procedure_T2}.
    \item We adopted a minimization procedure to extract the values of the coherence time and the detuning by matching the numerical (Fig. \ref{fig:apx_V2O3_ITRPP_2p4eV_procedure_T2}) and experimental (Fig. 3) values of $\Delta$ and $\delta \gamma/\gamma_p$. 
\end{itemize}

In the measurement performed at $T=\SI{100}{K}$, the energy position of the transition can be extracted also from the data displayed in the inset of Fig. 2b. By fitting a linear regression to the experimental data, we estimate $\hbar \omega_{01}=\left( \SI{2.37}{} \pm \SI{0.01}{} \right) \, \SI{}{eV}$. This value is comparable to $\left( \SI{2.39}{} \pm \SI{0.03}{} \right) \, \SI{}{eV}$ which is the value obtained by applying the minimization procedure described at the beginning of this Section.

\end{document}